\documentclass{article}

\usepackage[final]{nips_2018}

\usepackage[utf8]{inputenc} % allow utf-8 input
\usepackage[T1]{fontenc}    % use 8-bit T1 fonts
\usepackage{hyperref}       % hyperlinks
\usepackage{url}            % simple URL typesetting
\usepackage{booktabs}       % professional-quality tables
\usepackage{amsfonts}       % blackboard math symbols
\usepackage{nicefrac}       % compact symbols for 1/2, etc.
\usepackage{microtype}      % microtypography
\usepackage{graphicx}       % Required
\usepackage{amsmath}
\usepackage{multirow}
\usepackage{subcaption}

\title{Deep Multi-Agent Reinforcement Learning with Relevance Graphs}

\author{
  Aleksandra~Malysheva\thanks{Equal contribution.}\hspace{1.7mm}\thanks{Work performed while at Deep Learning Camp Jeju.}\\
  JetBrains Research\\
  National Research University \\ Higher School of Economics\\
  St. Petersburg, Russia \\
  \texttt{malysheva777@gmail.com} \\
  \And
  Tegg Taekyong~Sung\footnotemark[1]\hspace{1.7mm}\footnotemark[2] \\
  Department of Electronics and \\ Communications Engineering \\
  Kwangwoon University \\
  Seoul, Republic of Korea \\
  \texttt{tegg89@gmail.com} \\
  \AND
  Chae-Bong~Sohn\\
  Department of Electronics and \\ Communications Engineering \\
  Kwangwoon University \\
  Seoul, Republic of Korea \\
  \texttt{cbsohn@kw.ac.kr} \\
  \And
  Daniel~Kudenko \\
  University of York\\
  York, United Kingdom \\
  \texttt{daniel.kudenko@york.ac.uk} \\
  \AND
  Aleksei~Shpilman \\
  JetBrains Research\\
  National Research University \\ Higher School of Economics \\
  St. Petersburg, Russia \\
  \texttt{alexey@shpilman.com} \\
}

\begin{document}

\maketitle

\begin{abstract}
  Over recent years, deep reinforcement learning has shown strong successes in complex single-agent tasks, and more recently this approach has also been applied to multi-agent domains. In this paper, we propose a novel approach, called MAGnet, to multi-agent reinforcement learning (MARL) that utilizes a relevance graph representation of the environment obtained by a self-attention mechanism~\cite{vaswani2017attenion}, and a message-generation technique inspired by the NerveNet architecture ~\cite{wang2018nervenet}. We applied our MAGnet approach to the Pommerman game~\cite{matiisen2018pommerman} and the results show that it significantly outperforms state-of-the-art MARL solutions, including DQN, MADDPG, and MCTS. 
\end{abstract}

\section{Introduction}
\label{sec:introduction}
A common difficulty of reinforcement learning in a multi-agent environment is that in order to achieve successful coordination, agents require information about the relevance of environment objects to themselves and other agents. For example, in the game of Pommerman it is important to know how relevant bombs placed in the environment are for teammates, e.g. whether or not the bombs can threaten them. While such information can be hand-crafted into the state representation for well-understood environments, in lesser-known environments it is preferable to derive it as part of the learning process. 

In this paper, we propose a novel method, named MAGNet, to learn such relevance information in form of a relevance graph and incorporate this into the reinforcement learning process. Furthermore, we propose the use of message generation techniques from this graph, inspired by the NerveNet architecture~\cite{wang2018nervenet}. NerveNet has been introduced in the context of robot locomotion, where it has been applied to a graph of connected robot limbs. MAGNet uses a similar approach, but basing the message generation on the learned relevance graph.  

We applied MAGNet to the popular Pommerman~\cite{matiisen2018pommerman} multi-agent environment, and achieved significantly better performance than a baseline heuristic method and state-of-the-art RL techniques including DQN~\cite{mnih2013playing}, MADDPG~\cite{lowe2017MADDPG} and MCTS~\cite{guez2018learning}. Additionally, we empirically demonstrate the effectiveness of self-attention, graph sharing and message generating modules with an ablation study.

\section{Deep Multi-Agent Reinforcement Learning}
\label{sec:background}
In this section we describe the state-of-the-art (deep) reinforcement learning techniques that were applied to multi-agent domains. The algorithms introduced below (DQN, MCTSNet, and MADDPG) were also used as evaluation baselines in our experiments.

The majority of work in the area of reinforcement learning applies a Markov Decision Process (MDP) as a mathematical model~\cite{puterman2014markov}. An MDP is a tuple $(S,A,T,R)$, where $S$ is the state space, $A$ is the action space, $T(s,a,s')=Pr(s'|s,a)$ is the probability that action $a$ in state $s$ will lead to state $s'$,
and $R(s,a,s')$ is the immediate reward $r$ received when action $a$ taken in state $s$ results in a transition to state $s'$. The problem of solving an MDP is to find a policy (\textit{i.e.}, mapping from states to actions) which maximises the accumulated reward. When the environment dynamics (transition probabilities and reward function) are available, this task can be solved using policy iteration~\cite{bertsekas2005dynamic}. 

The problem of solving an multi-agent MDP is to find policies $\pi=\pi_1,\dots,\pi_N$ that maximize the expected reward $J_i=E_{s\sim p^\pi,a\sim\pi_i}[R]$ for every agent $i$, where $p^\pi$ is the distribution of states visited with policy $\pi$.

\subsection{Deep Q-Networks}
Q-learning is a value iteration method that tries to predict future rewards from current state and an action. This algorithm apply so called temporal-difference updates to propagate information about values of state-action pairs, $Q(s,a)$. After each transition, $(s,a)\rightarrow(s',r)$, in the environment, it updates state-action values by the formula:
\begin{align}\label{eq:1}
&    Q(s,a)\rightarrow Q(s,a)+\alpha[r+\gamma \max Q(s',a')-Q(s,a)], 
\end{align}
where $\alpha$ is the rate of learning and $\gamma$ is the discount factor. It modifies the value of taking action $a$ in state $s$, when after executing this action the environment returned reward $r$, and moved to a new state $s'$.

Deep Q-learning utilizes a neural network to predict Q-values of state-action pairs~\cite{mnih2013playing}. This so-called deep Q-network is trained to minimize the difference between predicted and actual Q-values as follows:
\begin{align}\label{eq:2}
&    y = r + \gamma \max_{a} Q^{past}(s',a')
\end{align}
\begin{align}\label{eq:3}
&    L(\theta) = \mathbb{E}_{s\sim \rho^\pi, a\sim p(s)}[y_i - Q(s,a|\theta)]^2,
\end{align}

where $y$ is best action according to the previous deep Q-network, $\theta$ is the parameter vector of the current Q-function and $a\sim p(s)$ denotes all actions that are permitted in state $s$.

The simplest way to apply this approach in multi-agent systems is to use an independent network for every agent~\cite{tan1993multi}. However this approach has been shown to not perform well with more complex environments~\cite{matignon2012independent}. One of the shortcomings of DQN learning in multi-agent settings is that past
experience replay is less informative, because unlike a single-agent setting, the same action in the same state may produce a different result based on the actions of other agents. A way to alleviate this problem is passing parameters of other agents as additional environmental information~\cite{tesauro2004extending}.

\subsection{Monte-Carlo Tree Search nets (MCTSNet)}
An alternative approach to reinforcement learning is to directly find an optimal policy without the intermediate step of computing a value function. Policy gradient methods (\textit{e.g.}~\cite{sutton2000policy}) have been developed to do just this.

Policy gradient methods have been shown to be successful in combination with Monte-Carlo tree search (MCTS)~\cite{chaslot2008monte}, which is a general, powerful, and widely used decision making algorithm, most commonly applied to games. In MCTS a sample tree of simulated future states is created, and evaluations of those states are backed-up to the root of this so-called search tree to compute the best action.

A recent study~\cite{guez2018learning} incorporates a neural network inside the tree-search by expanding, evaluating and backing-up a vector embedding of the states. The key idea is to assign a feature or a “memory” vector $h\in \mathbb{R}^n$ to an internal state (search tree node) that is then propagated up the tree and used to calculate the value or action in the root node. This MCTSNet approach has been shown to outperform other MCTS methods.

\subsection{Multi-agent Deep Deterministic Policy Gradient}
When dealing with continuous action spaces, the methods described above can not be applied. To overcome this limitation, the actor-critic approach to reinforcement learning was proposed~\cite{sutton2000policy}. In this approach an actor algorithm tries to output the best action vector and a critic tries to predict the value function for this action.

Specifically, in the Deep Deterministic Policy Gradient (DDPG~\cite{lillicrap2015DDPG}) algorithm two neural networks are used: $\mu(s)$ is the actor network that returns the action vector. $Q^w(s,a)$ is the critic network, that returns the $Q$ value, \textit{i.e.} the value estimate of the action of $a$ in state $s$. 

The gradient for the critic network can be calculated in the same way as the gradient for Deep Q-Networks described above (Equation~\ref{eq:3}). Knowing the critic gradient $\nabla_a Q^w$ we can then compute the gradient for the actor as follows:
\begin{align}
    \nabla_{\theta^\mu}J = \mathbb{E}_{s\sim \rho^\pi}[\nabla_a Q^w(s,a|\theta^w)|s=s_t,a=\mu(s|\theta^\mu)],
\end{align}
where $\theta^w$ and $\theta^\mu$ are parameters of critic and actor neural networks respectively, and $\rho^\pi(s)$ is the probability of reaching state $s$ with policy $\pi$.

The authors of~\cite{lowe2017MADDPG} proposed an extension of this method by creating multiple actors, each with it’s own critic with each critic taking in the respective agent’s observations and actions of all agents.

\section{MAGnet approach and architecture}
\label{sec:architecture}

\begin{figure*}[ht]
  \centering
    \includegraphics[height=0.65\linewidth]{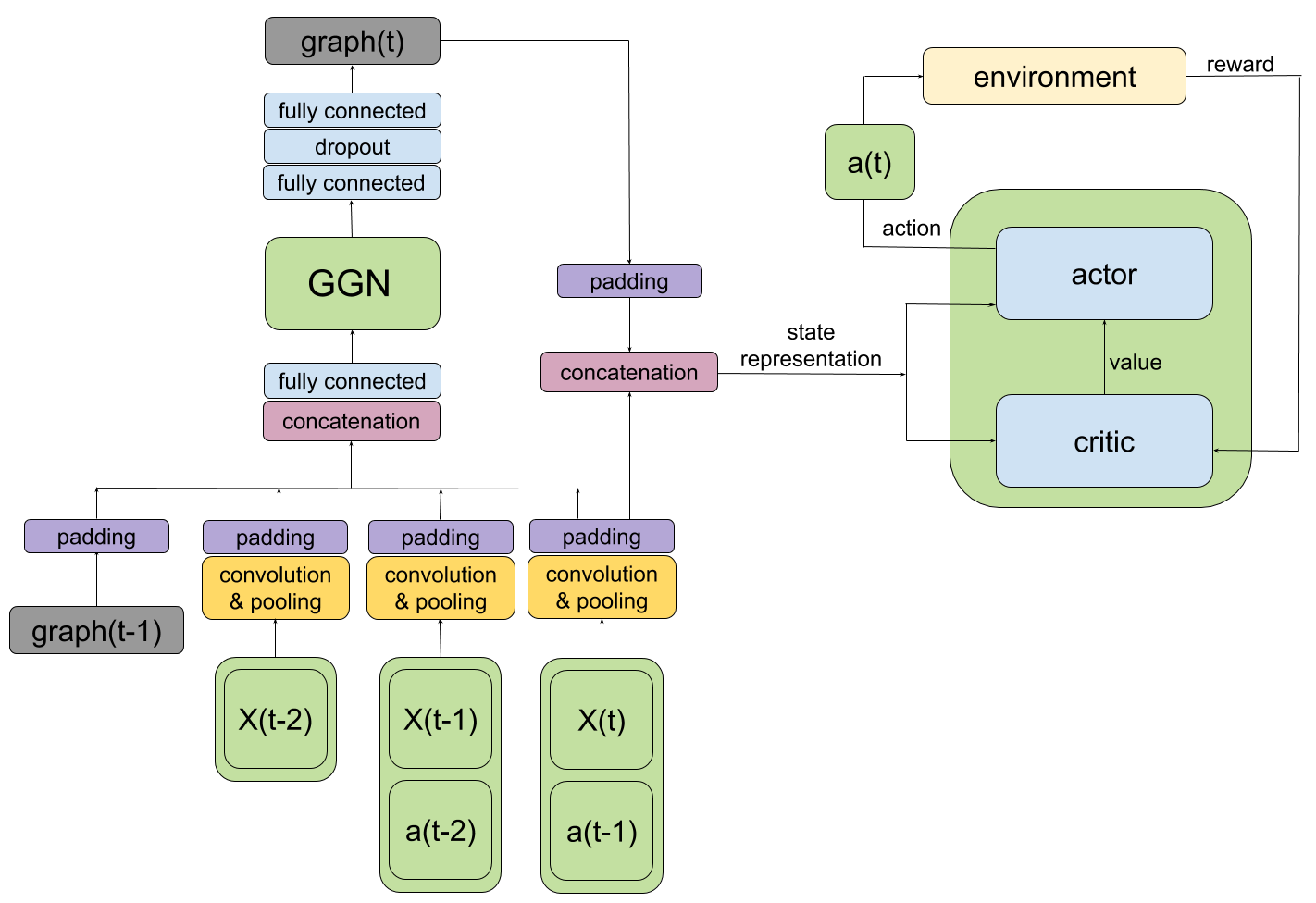}
    \caption{The overall network architecture of MAGNet. \textbf{Left} section shows the graph generation stage. \textbf{Right} part shows the decision making stage.}
    \label{img:curr-network}
\end{figure*}

The overall network architecture of our MAGNet approach is shown in Figure~\ref{img:curr-network}. The whole process can be divided into a relevance graph generation stage (shown in the left part) and a decision making stages (shown in the right part). We see them as a regression and classification problem respectively. In this architecture, the concatenation of the current state and previous action forms the input of the models, and the output is the next action. The details of the two processes are described below.

\subsection{Relevance graph generation stage}
In the first part of our MAGNet approach, a neural network is trained that produces a relevance graph. The relevance graph represents the relationship between agents and between agents and environment objects. The higher the weight of an edge between an agent $a$ and another agent $b$ or object $o$ is, the more important $b$ or $o$ are for the achievement of agent $a$'s task. The graph is generated by MAGNet from the current and previous state together with the respective actions.  

Figure~\ref{img:visual-graph}B shows an example of such a graph for two agents. The displayed graph only shows those edges which have a non-zero weight (thus there are objects to which agent 1 is not connected in the graph). 

In MAGNet, a neural network is trained via back-propagation to output a relevance graph represented as an $|A| \times (|A|+|O|)$ matrix, where $|A|$ is the number of agents and $|O|$ is the maximum number of environment objects. The input to the network are the current and the two previous states (denoted by $X(t)$, $X(t-1)$, and $X(t-2)$ in Figure~\ref{img:curr-network}), the two previous actions (denoted by $a(t-1)$ and $a(t-2)$), and the relevance graph produced at the previous time step (denoted by $graph(t-1)$). For the first learning step (i.e. $t=0$), the input consists out of three copies of the initial state, no actions, and a random relevance graph. The inputs are passed into a convolution and pooling layer, followed by a padding layer, and then concatenated and passed into fully connected layer and finally into the graph generation network (GGN). The GGN can be either a multilayer perceptron (MLP) or a self-attention network, which uses an attention mechanism to catch long and short term time-dependencies, and is an analogue to a recurrent network such as LSTM, but takes much less time to compute~\cite{vaswani2017attenion}. The result of the GGN is fed into a two-layer fully connected network with dropout, which produces the relevance graph matrix, as described above.

The loss function for the back-propagation training is composed of two parts:
\begin{align}\label{eq:5}
    L = \|W_t-W_{t-1}\|^2_2 + \sum_{\xi\sim\Xi}(w_t(\xi) - s(\xi))^2
\end{align}

The first component is based on the difference between the current graph $W_t$ and the one generated in the previous state $W_{t_1}$. It is important to note that graph on each step is the same only weights are changing. The second component comes into play when a special pre-defined event $\xi\in\Xi$ occurs, and is based on the difference between a selected edge weight updated according to heuristic rules $s(\xi)$ and the weight of the same edge in the current graph $w_t(\xi)$.  For example, a heuristic rule would specify that if a bomb explodes and kills the agent, the edge weight between the agent and the bomb is set to a high value (i.e. the bomb is clearly of high relevance of the agent).

The training of the neural network can be performed in two stages: first with a default rule-based AI agent, and then with a learning agent. 

\subsection{Decision making stage}
The agent AI responsible for decision making is also represented as a neural network whose inputs are accumulated messages (generated by a method inspired by NerveNet~\cite{wang2018nervenet} and described below) and the current state. The output of the network is an action to be executed. 

The graph $G$ generated at the last step is $G = (V, E)$ where edges represent relevance between agents and objects. Every vertex $v$ has a type: $b(v) \in \{0, 1, 2, 3, 4, 5, 6\}$ that in our case corresponds to: "ally", "enemy", "placed bomb" (about to explode), "increase kick ability", "increase blast power", "extra bomb" (can be picked up). Every edge has a type as well: $c(e) \in \{0, 1\}$, that corresponds to “edge between the agents” and “edge between the agent and the object in the environment”.

The final (action) vector is computed in 4 stages through message passing system, similar to a system used for distributed computing and described in~\cite{attiya2004distributed}. Stages 2 and 3 are repeated for a specified number of message propagation steps.
\begin{enumerate}
    \item \textbf{Initialization of information vector.} Each vertex $v$ has an initialization network $MLP^{b(v)}_{init}$ associated with it according to it’s type $b(v)$ that takes as input the current individual observation $O_v$ and outputs initial information vector $\mu^0_v$ for each vertex.
    \begin{equation}
        \mu^0_v = MLP^{b(v)}_{init}(O_v)
    \end{equation}
    \item \textbf{Message generation.} At message propagation step $t+1$ message networks $MLP^{c(v,u)}_{mess}$ compute output messages for every edge $(v, u) \in E$ based on type of the edge $c(v, u)$.
    \begin{equation}
        m^t_{(v,u)} = MLP^{c(v,u)}_m (\mu^t_v)
    \end{equation}
    \item \textbf{Message processing.} Information vector $m^{t+1}_v$ at message propagation step $t$ is updated by update network $LSTM^{b(v)}_{up}$ associated with it according to it’s type $b(v)$, that takes as input a sum of all message vectors from connected to $v$ edges multiplied by the edge relevance $w_{(v,\ast)}$ and information at previous step $m^t_v$.
    \begin{equation}
        \mu^{t+1}_v = LSTM^{b(v)}_{up} (\mu^t_v, \sum m^t_{(v,\ast)}w_{(v,\ast)})
    \end{equation}
    \item \textbf{Choice of action.} All vertices that are associated with agents have a decision network $MLP^{b(v)}_{choice}$ which takes as an input its final information vector $m^t_v$ and compute the mean of the action of the Gaussian policy.
    \begin{equation}
        a_v = MLP^{b(v)}_{choice}(\mu^t_v)
    \end{equation}
\end{enumerate}

All networks are trained using back-propagation following the DDPG actor-critic approach~\cite{lillicrap2015DDPG}.

\section{Experiments}
\label{sec:experiments}

\subsection{Environment}
In this paper, we use popular Pommerman game environment which can be played by up to 4 players~\cite{matiisen2018pommerman}. This game has been used in many empirical evaluations of multi-agent algorithms, and therefore is especially suitable for a comparison to state-of-the-art techniques. In Pommerman, the environment is a grid-world where each agent can move in one of four directions, lay a bomb, or do nothing. A grid square is either clear (which means that an agent can enter it), wooden, or rigid. Wooden grid squares can not be entered, but can be destroyed by a bomb (i.e. turned into clear squares). Rigid squares are indestructible and impassable. When a wooden square is destroyed, there is a probability of items appearing, e.g., an extra bomb, a bomb range increase, or a kick ability. Once a bomb has been placed in a grid square it explodes after 10 time steps. The explosion destroys any wooden square within range 1 and kills any agent within range 4. The last surviving agent wins the gameast 

The map of the environment is randomly generated for every episode. The game has two different modes: free for all and team match. Our experiments were carried out in the team match mode in order to evaluate the ability of MAGnet to exploit the discovered relationships between agents (e.g. being on the same team).

\subsection{Network training}
We first trained the graph generating network on 50,000 episodes with the default Pommerman AI as the decision making agent. After this initial training, the default AI was replaced with the learning decision making AI described in section~\ref{sec:architecture}. All learning graphs show the training episodes starting with this replacement (except the ones which explicitly show the relevance graph learning).

Table~\ref{tbl:modules} shows results for different MAGNet variants in terms of achieved win percentage against a default agent after 50,000 episodes. The MAGNet variants are differing in the complexity of the approach, starting from the simplest version which takes the relevance graph as a direct input, to the version incorporating message generation, graph sharing, and self-attention. The table clearly shows the benefit of each extension.

\begin{table}[ht]
  \caption{Influence of different modules on the performance of the MAGnet model.}
  \label{tbl:modules}
  \begin{center}
  \begin{tabular}{| p{1.5cm} | p{1.5cm} | p{1.8cm} | p{1.6cm} |}
    \hline
    \multicolumn{3}{| c |}{\textbf{MAGnet modules}}  & \textbf{Win \%}\\ 
    \cline{1-3}
    \textbf{Self-attention} & \textbf{Graph Sharing} & \textbf{Message Generation} & \\
    \hline
    + & + & + & \boldmath$71.3\pm 0.7$\\ \hline
    + & + & - & $56.7\pm 1.8$\\ \hline
    + & - & + & $62.4\pm 1.7$\\ \hline
    + & - & - & $54.5\pm 2.6$\\ \hline
    - & + & + & $67.1\pm 1.9$\\ \hline
    - & + & - & $52.0\pm 1.7$\\ \hline
    - & - & + & $45.2\pm 3.6$\\ \hline
    - & - & - & $32.7\pm 5.9$\\ \hline
  \end{tabular}
  \end{center}
\end{table}

Each of the three extensions with their hyper-parameters are described below: 

Graph Generating Network (\textbf{GGN}): we used a MLP (number of layers and neurons was varied, and a network with 3 layers 512-128-128 neurons achieved the best result) and a self-attention (\textbf{SA}) layer~\cite{vaswani2017attenion} with default parameters.

Graph Sharing (\textbf{GS}): relevance graphs were trained individually for both agents, or in form of a shared graph for both agents.

Message Generation (\textbf{MG}): the message generation module was implemented as either a MLP or a message generation (MG) architecture, as described in Section~\ref{sec:architecture}. We tested the MLP and message generation network with a range of hyper-parameters. For the MLP with 3 fully connected layers 1024-256-64 neurons achieved the best result, while for the message generation network 2 layers with 128-32 neurons and 5 message passing iterations showed the best result.

Dropout layers were individually optimized by grid search in [0, 0.2, 0.4] space.

We tested two convolution sized: [3x3] and [5x5]. [5x5] convolutions showed the best result.

Rectified Linear Unit (ReLU) transformation was used for all connections.

\subsection{Evaluation Baselines}
\begin{figure}[!ht]
  \centering
  \begin{subfigure}[!ht]{0.45\linewidth}
    \centering
    \includegraphics[width=\linewidth]{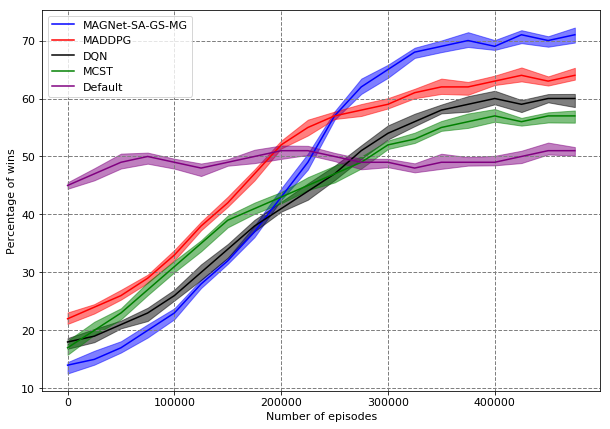}
    \caption{}
    \label{img:baselines}
  \end{subfigure}
  ~
  \begin{subfigure}[!ht]{0.45\linewidth}
   \centering
    \includegraphics[width=\linewidth]{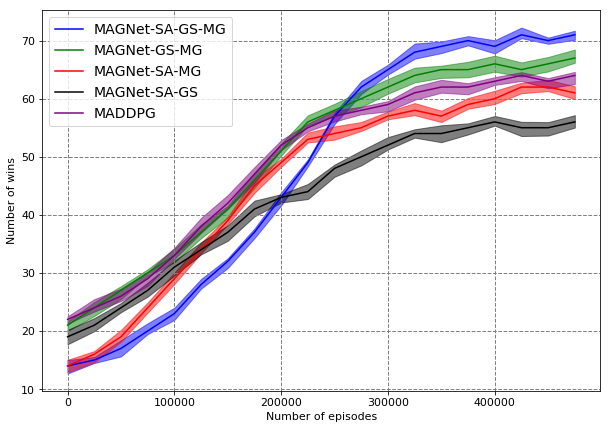}
    \caption{}
  \end{subfigure}
  \caption{(a) The best performing MAGnet variant (MAGnet-Att-NerveNet-GS) compared to state- of-the-art MARL techniques. (b) The effectiveness MAGNet with various module combinations: Message Generation (MG), shared relevance graph (GS), and Self-Attention (SA). MADDPG is currently the best performing state-of-the-art algorithm.}
  \label{img:best}
  \vspace{-3mm}
\end{figure}

In our experiments, we compare the proposed method with state-of-the-art reinforcement learning algorithms simulated in team match mode. Figure~\ref{img:baselines} shows a comparison with DQN~\cite{mnih2013playing}, MCTSNets~\cite{guez2018learning}, MADDPG\cite{lowe2017MADDPG}, and a default heuristic AI. The latter algorithm is provided as part of the Pommerman environment~\cite{matiisen2018pommerman}. Each of the reinforcment learning algoritms played a number of games (i.e. episodes) against the heuristic AI, and the respective win rates are shown.

All graphs display a 95\% confidence interval to illustrate the statistical significance of our results.

The parameters chosen for the baselines were set as follows.

For \textbf{DQN} we implemented multi-agent deep Q-learning approach which has been shown to be successful in past work~\cite{egorov2016multi}. In this method training is performed in two repeated steps: first, one agent is training at a time, while policies of other agents are kept fixed; second, the agent that was trained in the previous step distributes its policy to all of its allies as an additional environmental variable.

The network consists of five convolutional layers with 64 3x3 filters in each layer followed by three fully connected layers with 128 neurons each with residual connections~\cite{he2016deep} and batch normalization~\cite{ioffe2015batch} that takes an input an 11x11x4 environment tensor and one-hot encoded action vector (a padded 1x6 vector) that are provided by the Pommerman environment and outputs a Q-function for that state. This network showed the best result at the parameter exploration stage.

Parameter exploration on \textbf{MCTSNet} led to the following settings: \textbf{The backup network} $\beta$ is a multilayer perceptron (MLP) with 5 fully connected layer with 64 neurons in each layer that takes in current "memory" vectors of the node and updated "memory" vector of the child and updated the node’s "memory" vector. \textbf{The embedding network} $\epsilon$, is consists of 7 convolutional layers with 64 3x3 filters followed by 3 fully connected layers with 128 neurons each with residual connections~\cite{he2016deep} and batch normalization~\cite{ioffe2015batch} that takes an input an 11x11x4 environment tensor and one-hot encoded action vector (a padded 1x6 vector) that are provided by Pommerman and outputs a "memory" vector. \textbf{The policy network} has the same architecture, but with 5 convolutional layers with 32 3x3 filters each and it outputs an action for simulation. \textbf{The readout network}, is a multilayer perceptron with 2 fully connected layer with 128 neurons in each layer that inputs root "memory" vector and outputs an action.

For our implementation of \textbf{MADDPG} we used a multilayer perceptron (MLP) with 5 fully connected layer with 128 neurons in each layer and for the critic we used a 3 layer network with 128 neurons in each layer.

\subsection{Self-attention and graph sharing in training a relevance graph}
\begin{figure}[!ht]
  \centering
  \begin{subfigure}[!ht]{0.45\linewidth}
    \centering
    \includegraphics[width=\linewidth]{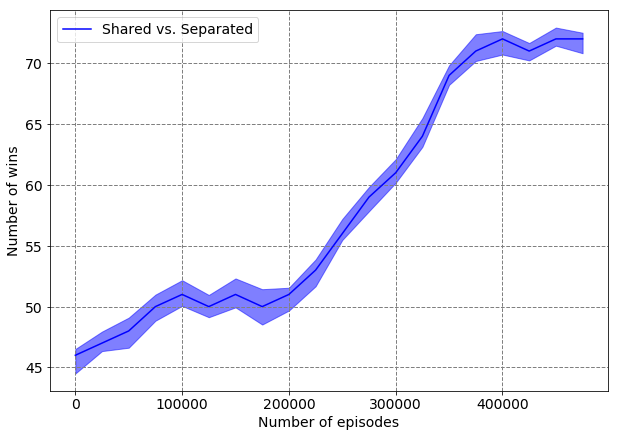}
    \caption{}
    \label{img:graph-evaluation}
  \end{subfigure}
  ~
  \begin{subfigure}[!ht]{0.45\linewidth}
   \centering
    \includegraphics[width=\linewidth]{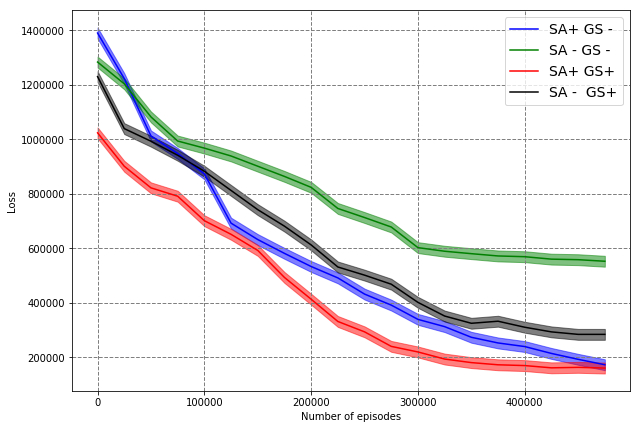}
    \caption{}
  \end{subfigure}
  \caption{(a) Win rate of a MAGNet team with a shared relevance graph vs a MAGNet team with agent-individual relevance graphs. (b) Loss value in training the graph generator with and without a self-attention module (SA+/-) and with or without graph sharing (GS+/-).}
  \label{img:attention-exp}
  \vspace{-3mm}
\end{figure}

Figure~\ref{img:attention-exp} shows the shared graph loss value (Equation~\ref{eq:5}) with and without self-attention module and with or without graph sharing. As we can see from this figure, both self-attention and graph sharing significantly improve graph generation in terms of speed of convergence and final loss value. Furthermore, their actions are somewhat independent which is seen in that using them together gives additional improvement.

To provide further evidence for the usefulness of the shared graph approach, we let a MAGNet-AttNerveNet team play against a MAGNet-Att-NerveNet-GS team. As the graph in~\ref{img:graph-evaluation} shows, even though both have the same base architectures, the graph sharing method yields a higher win-rate after 10,000 episodes.

\subsection{Relevance graph visualization}
Figure~\ref{img:visual-graph} shows examples of relevance graphs with the corresponding environment states. Red nodes denote friendly team agents, the purple nodes denote the agents on the opposing team, and the other nodes denote environment objects such as walls (green) and bombs (black). The lengths of edges represent their weights (shorter edge equals higher weight, i.e. higher relevance). The graphs in Figure~\ref{img:visual-graph}B are shared, while the graphs in Figure~\ref{img:visual-graph}C are agent-individual.

As can be seen when comparing the individual and shared graphs, in the shared case agent 1 and agent 2 have different strategies related to the opponent agents (agents 3 and 4). Agent 4 is of relevance to agent 1 but not to agent 2. Similarly, agent 3 is of relevance to agent 2, but not to agent 1. In contrast, when considering the individual graphs, both agents 3 and 4 have the same relevance to agents 1 and 2. Furthermore, it can be seen from all graphs that different environment objects are relevant to different agents. 

\section{Conclusion}
\label{sec:conclusion}
In this paper we presented a novel method, MAGNet, for deep multi-agent reinforcement learning incorporating information on the relevance of other agents and environment objects to the RL agent. We also extended this basic approach with various optimizations, namely self-attention, shared relevance graphs, and message generation inspired by Nervenet. The MAGNet variants were evaluated on the popular Pommerman game environment, and compared to state-of-the-art MARL techniques. Our results show that MAGNet significantly outperforms all competitors. 

% The source code used to train and evaluate our models is available at \url{https://github.com/tegg89/magnet}.

\section{Acknowledgments}
This was supported by Deep Learning Camp Jeju 2018 which was organized by TensorFlow Korea User Group. This work was partially supported by the Ministry of Trade, Industry \& Energy (MOTIE, Korea) under Industrial Technology Innovation Program No.10077659, `Development of artificial intelligence based mobile manipulator for automation of logistics in manufacturing line and logistics center'.

\newpage
\bibliographystyle{abbrv}
\bibliography{references}

\newpage
\appendix
\section{Visualization of Relevance Graph}
\begin{figure*}[htb]
  \centering
    \includegraphics[width=1.0\linewidth]{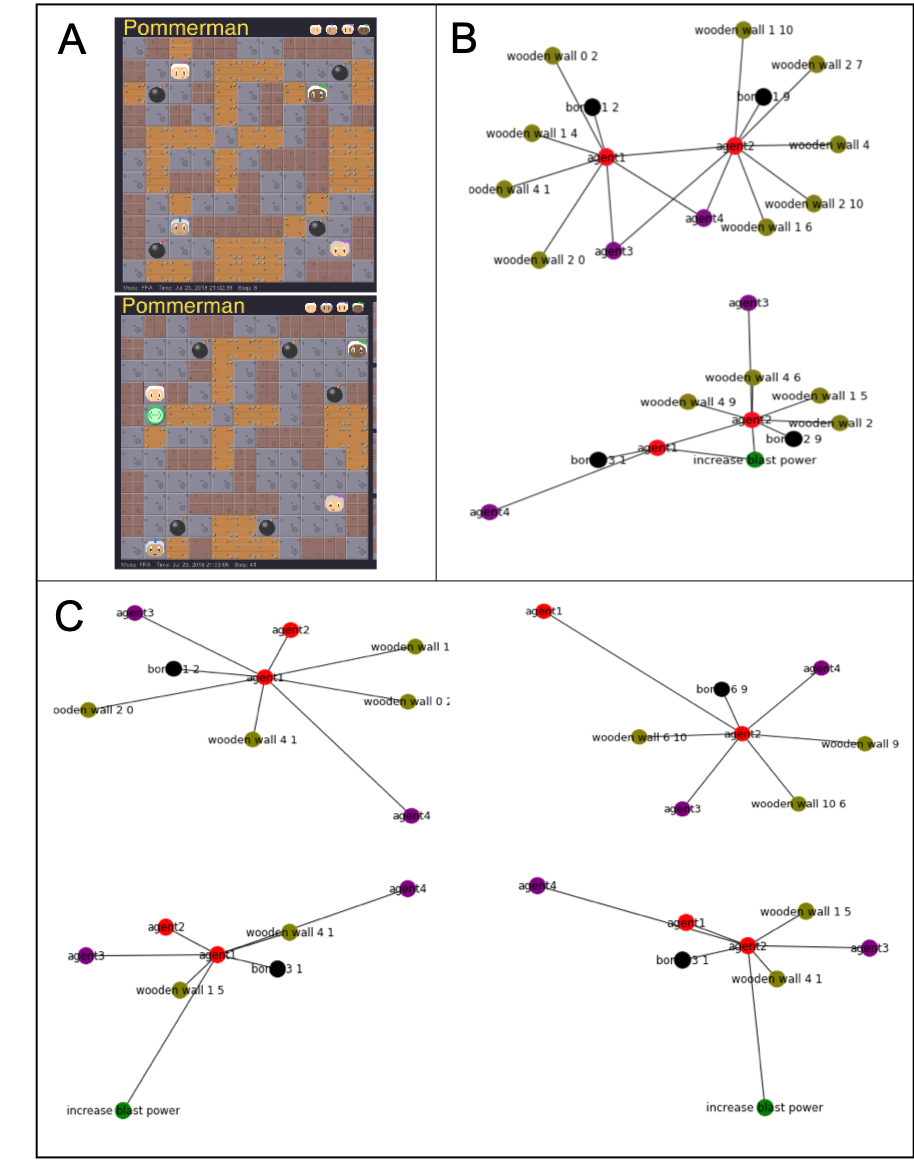}
    \caption{Visualization of relevance graph. (A) Corresponding game states. (B) Shared graph. (C) Agent-individual graphs.}
    \label{img:visual-graph}
\end{figure*}

\end{document}